\documentstyle[amssymb,multicol,epsf,aps]{revtex}
\def\dbarrm {{\mathchar'26\mkern-11mu{\rm d}}}                         %
\newcommand{\BEQ}{\begin{equation}}
\newcommand{\EEQ}{\end{equation}}
\newcommand{\BEA}{\begin{eqnarray}}
\newcommand{\EEA}{\end{eqnarray}}
\renewcommand{\H}{{\cal H}}

\renewcommand{\d}{{\rm d }}

\newcommand{\I}{{\cal I}}
\newcommand{\p}{\partial}
\renewcommand{\S}{S_{\rm ep}}

\begin{document}
\draft
\title{Formulation of thermodynamics for the glassy state:\\
configurational energy as a modest source of energy}
\author{Th.M. ~ Nieuwenhuizen}
\address{Department of Physics and Astronomy, University of Amsterdam
\\ Valckenierstraat 65, 1018 XE Amsterdam, The Netherlands}
\maketitle
\begin{abstract}
Glass is an under-cooled  liquid that very slowly relaxes  towards
the equilibrium crystalline  state. Its energy  balance is ill understood,
since it is widely believed that the glassy state cannot  
be described thermodynamically.
However, the classical paradoxes involving the Ehrenfest relations 
and Prigogine-Defay ratio
can be explained when the `effective' or `fictive' temperature
of the slow modes is taken as additional system parameter.

Having straigtened out the proper picture, it is interesting
to reconsider glass from a thermodynamic viewpoint.
A shard of glass, kept at fixed temperature and volume, relaxes  
towards lower energy. Heat is released,
inducing apparent violations of all basic thermodynamic laws.
The most interesting application is to use glassy or amorphous
systems as a source of energy, by extracting the configurational energy 
in a process of melting followed by crystallization.
\end{abstract}

\begin{multicols}{2}

Glasses have been of use to mankind  from early  on, be it as arrowheads
for the stone age people of Corsica and the Americas, the obsidian battle
axes and swords of the Aztecs,  pumice scrappers for animal hides, or the
tektite  ornaments and fertility symbols of our ancestors. It has also been
a source of fascination once its manufacturing process became understood
around 2000BC in the Caucasus   and has continued to amaze  us when
considering artifacts from the tomb of Thutmose III(1480 BC), stained
glasses windows in Chartres or the exciting world of silica glass based
fiber optics.

Besides silica based ones, there exist glasses such as borate, phosphate, or
germanate glasses,  which have as primary and sufficient attribute the four
fold oxygen coordination of the glass forming cation. Glassy ice
can exist as crystals  in forms topologically equivalent to silica
polymorphs. Also oxygen free glasses as the chalcogenide  or halide
ones exists, as do polymer glasses which go under the house hold names of
nylon, PVC bottles, or handywrap. For more examples, see~\cite{deJong}.

The archetypical glass for most of us is window glass, a fairly late
invention around 500 AD, which commonly consists for about 70 percent of
silica, the remainder  being partitioned  in about  equal proportions
between soda and lime. It forms films which flow slowly causing a thinning
at the upper end in the course of time, and time lapse pictures taken every
500 years would reveal that  its behavior would be akin to that of a soap
film.

What all these glassy systems have in common  is that they will
structurally relax from an excited glassy state to a crystalline ground
state in due course. Such relaxation may be slow as is typical for systems
with broken ergodicity, i.e. systems which do not reach equilibrium within
a characteristic observation time. 

It is the nature of this structural relaxation which we wish to 
consider here in the light of a recently developed 
thermodynamic structure for glassy systems. 
After doing that, we shall reconsider the glass from a thermodynamic
view point, and find apparant violations of all basic laws.

\subsection*{The glassy state: a diverging time scale}

Any liquid cooled sufficiently fast will become glassy, i.e. it will lack 
the time to evolve into a long-range ordered  crystalline array. 
In such glassy materials two types of processes can be distinguished: 
fast or $\beta$ processes, which happen on short scale and
remain in thermal equilibrium, and slow or $\alpha$-processes, 
which are restructurization processes that involve many atoms. 
Somewhere during cooling they fall out of thermal equilibrium, or,
more precisely, their relaxation time exceeds the time needed to
reach the (sliding) equilibrium. Recently the microscopic nature of
the liquid landscape that leads to these slow processes was discussed 
~\cite{AAngel}\cite{Shastry}.

The relaxation time  in glassy systems scales with the viscosity, which is
a measure of the internal friction of a system. It has in general a strong
temperature dependence varying in silica-based glasses commonly  over 18
orders of magnitude  in $P\!a.s$, and is the least understood of all glass
properties. Both viscosity and relaxation  time will grow sufficiently
rapid in a typical cooling experiment that a point is reached  where 
equilibrium no longer exists between the thermal state of the system and the
surrounding temperature.  This point is called the glass transition
temperature,  commonly  occuring in the temperature range of roughly  two
thirds of the melting temperature in silica based glasses.

The growth in relaxation time  for so-called fragile glasses,
such as toluene, glycerol, and ethanol,  is built into
the Vogel-Fulcher-Tammann-Hesse law $\tau_{\rm eq}\sim \exp[B/(T-T_0)]$
\cite{Vogel}\cite{Fulcher} \cite{Tamman}.
Adam and Gibbs~\cite{AdamGibbs} provided an intuitive explanation 
of this law and Parisi~\cite{Parisiinstanton} gave a field
theoretical argument for the related behavior
$\tau_{\rm eq}\sim \exp\{[B/(T-T_0)]^{d-1}\}$, where $d$ is the 
spatial dimension of the system  ($d=3$ in practice).
For so-called  strong glass, such as SiO$_2$ and  GeO$_2$, 
on the other hand, the
relaxation time follows an Arhenius law $\tau_\alpha\sim \exp(A/T)$. 
Somewhat related expressions were proposed 
by  Kivelson et al. ~\cite{Kivelson} and by Schultz ~\cite{MSchultz}
to describe  relaxation times for all known glasses todate.

In our studies we have benefited from exactly solving simple model systems 
with an Arrhenius law~\cite{BPR}\cite{Nhammer}\cite{Nlongthermo}. 
or a VFTH-law~\cite{NVFmodel,LN}. 
In models such as soft-spin models ~\cite{KT}, the 
spherical $p$-spin model~\cite{CHS}, \cite{Nthermo}, 
or a directed polymer model~\cite{Ndirpol} partial answers can be obtained.
Together this has resulted in a coherent picture that we
 discuss in the following.

\subsection*{Breakdown of thermodynamics  for the glassy state?}

Attempts for a thermodynamic interpretation of  a glass phase go back
to the twenties. Let us consider the experimental way to determine
the entropy: one integrates $\d S=(C_p/T)\d T$, 
where $C_p$ is the measured heat capacity, from a sufficiently high 
reference temperature in the liquid phase, down to lower
 temperatures. The excess entropy is the thus obtained entropy minus
the entropy of the crystalline state.
It decreases with temperature, but will remain
positive down to zero temperature, thus resulting in a
residual entropy, in apparent violation with the third law.

The glass transition temperature depends on the cooling rate.
Slower cooling  provides the system with more time to
relax, with as result that it will remain in equilibrium  to
lower temperatures.  Kauzmann made it plausible that,
for infinitely slow cooling, $\Delta S$ goes to zero near a specific
temperature, now called the Kauzmann temperature $T_K$~\cite{Kauzmann}.
It is to be expected that $T_K$ and $T_0$ coincide, and this indeed
occurs in an exactly solvable model glass~\cite{NVFmodel,LN}.
The ultimate vanishing of the excess
entropy near this point is called the {\it Kauzmann paradox}.
A thermodynamic Kauzmann transition
has been found in mean field models, like the
Random Energy Model~\cite{Derrida}, the spherical $p$-spin model
~\cite{CSstat},
and a polymer model~\cite{Ndirpol}, but also in a model
obeying a VFTH-law~\cite{NVFmodel,LN}. Finally, it also shows up
in a mean field  approach (so-called replica approach)
to the glass transition in realistic models for glass~\cite{MP}.

Proposals to formulate properties of the glassy state in
a thermodynamic frame have remained unsuccessful till recently
when we adressed this issue.
The negative conclusions were not based on subtle entropy arguments,
but on  testing the so-called Ehrenfest relations (see the Appendix).
They are relations between the smeared out discontinuities that
occur around the glass formation temperature in cooling experiments 
at constant pressure,  in the specific
heat $C_p=\p(U+pV)/\p T|_p$ and the expansivity
$\alpha=\p\ln V/\p T|_p$.
One can treat them as true discontinuities, because of the small size of the
smearing region. When comparing cooling experiments at different pressures
one also finds discontinuities in the compresibility
$\kappa=-\p\ln V/\p p|_T$. A glass transition seems therefore to be
a textbook  example of a mean field second order phase transition.
However, this analogy is not perfect because  there is no diverging length
scale, and the smaller value of the specific heat occurs in
the low temperature regime.

The two Ehrenfest relations are particularly useful
tests for glasses. The common notion of half a century of research 
is that the first relation is always violated, whereas
 the second relation is obeyed in most cases,
though not in all~\cite{McKenna}\cite{Angell}.
In the most careful experiment we came across, the glass transition in
atactic polystyrene with a cooling rate of $18\,K/h$ at ambient pressure,
$\Delta C_p$ is too small by a factor $0.77$
\cite{RehageOels}, while $\Pi=1.09\approx 1.$

The failure of in particular the first Ehrenfest
relations has led to the general belief that {\it ``thermodynamics
 does not work for glasses, as there is no
equilibrium''}. This unapplicability to glasses is unsatisfactory on two
accounts. Firstly, thermodynamics should also apply to non-equilibrium
systems. Secondly, in view of the large time scales, up to many years,
one would hope for a universal description of the energy balance.

The negative conclusion about the applicability of thermodynamics
was mainly based on the failure to understand the Ehrenfest relations 
and the Prigogine-Defay ratio.  It should be kept in mind
that, so far, the approaches leaned very much on equilibrium ideas.
Well known examples are the 1953 Davies-Jones theory~\cite{DaviesJones},
the 1958 Gibbs-DiMarzio
{}~\cite{GibbsDiMarzio} and the 1965 Adam-Gibbs~\cite{AdamGibbs} papers,
while a 1981 paper by DiMarzio has title ``Equilibrium theory of
glasses'' and a subsection ``An equilibrium theory of glasses is
absolutely necessary''~\cite{DiMarzio1981}.
In particular, the Gibbs-DiMarzio model has long been considered
as an important theory for glass.
It is our purpose to explain that the step to an 
equilibrium theory has been an unfortunate move.

\subsection*{Thermodynamics far from equilibrium and 
the effective or fictive temperature}

Though a glass may reach equilibrium after many thousands of years, it
will be glassy on the timescale of a few years. In such non-equilibrium
systems the elapsed time $t$ , sometimes called the {\it observation time},
{\it waiting time}, or {\it age}, is an important variable
for the description of the macroscopic state of the system. 
In realistic glass forming liquids several 
types of slow processes may occur. 
Here  we shall  restrict ourselves to 
model glasses with one type of slow process, 
because this already explains the basic problem.
For such systems we have found that time shows up through 
one or two time-dependent parameters, namely the effective temperature
 $T_e(t)$~\cite{NEhren}, and possibly the effective 
pressure $p_e(t)$~\cite{NVFmodel}.

$T_e$ was introduced phenomenologically half a century ago 
by Tool~\cite{Tool} to model the shapes of the specific heat 
curves on cooling and heating through the glass transition.
This effective or fitctive temperature describes the relaxation of slow modes. 
As in a plasma, slow and fast modes will equilibrate
at their own specific temperature. 
We have recently proposed a set of non-linear cooling and heating
experiments, that should lead to system-independent shapes of the specific
heat near the glass transition~\cite{Nhammer}\cite{Nlongthermo}.

The notion of an effective pressure in the glassy state
was put forward by Goldstein
~\cite{Goldstein} and J\"ackle~\cite{Jaeckle}. 
They considered glass formation under pressure. If pressure is partly
released after the glass has formed, the glassy state will depend
not so strong on the new pressure, but more on its formation pressure, 
which thus acts as a fictive pressure.  In an exactly solvable
model glass we found a range of parameters where such a parameter has to be
introduced for a proper thermodynamic description~\cite{NVFmodel,LN}.
The approach with two effective variables is promising, since
for glass forming liquids it was found experimentally that 
at given $T,p$ the same volume can be reached via different histories.
As these histories then lead to different futures,  the glassy 
state of these systems cannot be coded using only one 
extra variable~\cite{McKenna}.
Here we shall, however,
 avoid the technical complications of having to deal with
two effective parameters.

Let us explain how the effective temperature arises in our approach.
Consider an aging experiment at fixed $(T,p)$ at  
some long observation time $t$. 
Fast modes, with relaxation time much less than $t$,
will be in equilibrium with the bath temperature $T$.
Modes with timescale much larger than $t$ are  just quenched.
The active slow modes are those that have a time scale comparable with $t$. 
For dynamics on this timescale one may sum out the fast processes, 
by calculating the restricted partition sum, i.e.
the sum over the fast modes of the Boltzmann weights
$\exp(-\H/kT)$, where $\H$ is the Hamiltonian, i.e. the energy of the
configuration, and $k$ Boltzmann's constant. 
We have worked this out in detail for a model system~\cite{NVFmodel,LN}, 
but here we donot restrict ourselves to a specific system.
(How to divide between fast and slow modes in realistic
systems is a subject of current research). 
The partition sum leads to an effective Hamiltonian of the 
active slow modes, 
$\H_{\rm eff}({\rm active})=U-T\S$ where $U$ 
is the energy, averaged over the equilibrium processes, 
and $\S$ their entropy.  
$U$ and $\S$ depend explicitly on the actual 
configuration of active slow modes, 
but also on $T$, which results from the fast modes.
The effective Hamiltonian governs the active slow dynamics, yielding
the dynamical solution for $U(t)$.
One may describe this result by quasi-static expressions
that follow from a partition sum at an effective temperature 
$T_e$. It is introduced by now taking as Boltzmann factor: 
$\exp(-\H_{\rm eff}/kT_e)$.
The sum over slow modes will yield a generalized free  enthalpy
\BEQ \label{Ge=} G(T,T_e,p)=U-T\S-T_e\I+pV \EEQ			
now $U$ and $\S$ have been averaged over the active slow modes, and
$\I$ is their {\it configurational entropy}. 
It is also called {\it information entropy} or {\it complexity}.
 $T_e(t; T,p)$ is now fixed by matching $U(T,T_e,p)$
with the  dynamical value $U(t;T,p)$.

In an aging experiment at fixed $(T,p)$ both $T_e$ will be a function
of time. In a cooling experiment at fixed pressure, $T$ and $T_e$
will be functions  of time, so that a line in the $(T,T_e,p)$ 
space is singled out. For a set of
smoothly related cooling experiments (e.g. having common cooling rates) at
different pressures, this leads to thermodynamics
confined to a surface $T_e(T,p)$ in ($T$, $T_e$, $p$) 
space~\cite{NEhren}.
To cover full space many sets of cooling experiments would be needed, e.g.
at different cooling rates. The results should coincide with 
other measurements, such as heating, compression and aging experiments.

Very recently numerical data of the (short-time) vibrational
properties in the glassy phase of a binary Lennard-Jones system 
were observed to be governed by an effective temperature, 
in full harmony with our picture~\cite{KobSciorTart}.

\subsubsection*{The first Ehrenfest relation}

The two-temperature approach explains in one stroke the confusion 
about the first Ehrenfest relation, eq. (\ref{Ehren1}), 
believed to be {\it never satisfied}. 
The misunderstanding arose because
in the glassy state the compressibility is not a unique material
parameter: this measure of the system's response to a change in
pressure depends on how the measurement is done, and basically on how
much time the system is given.
This fact has already severely complicated its experimental determination
 and interpretation.  In the past, one has typically obtained 
$\alpha$ and $\d p_g/\d T$ from the cooling curves at fixed pressures, 
but determined $\kappa$ by an alternative experiment, such as the 
speed of sound, a procedure which normally works satisfactorily. 
In retrospective, this unvalid approach can be traced back to a far 
too large trust in theories which view glass
transitions as thermodynamic phase transitions~\cite{PrigogineDefay}
\cite{DaviesJones}\cite{GibbsDiMarzio}\cite{AdamGibbs}\cite{Moynihan}.
In such a description a unique long-time limit of the compressibility exists,
that should be measurable in various ways.
McKenna has stressed that in experiments on glasses the 
isothermal compressibility differs from the isochoral compressibility
~\cite{McKenna}. Thus alternative experiments are not allowed, and
there remains only a tautology to verify. The physical reason is that 
the compressibility strongly depends on the amount of time 
(and other details of the history) that the system has to relax.
This effect immediately sets in below the glass transition. 
In our approach the validity of the first Ehrenfest relation 
can be checked explicitly~\cite{NEhren}.

\subsubsection*{The second law and the second Ehrenfest relation}

Thermodynamics amounts to giving universal relations for the system's 
state variables at nearby points in  ($T$, $T_e$, $p$) space.
The property $\d G=-\S\d T-\I\d T_e+V\d p$ implies together with 
eq. (\ref{Ge=}) and the first law 
$\d U=\dbarrm Q+\dbarrm W=\dbarrm Q-p\d V$, that the supplied heat is
\BEQ\label{2law}
\dbarrm Q=T\d \S+T_e\d\I\EEQ
The entropy terms are the same as for a cup of coffee in a room:
also then there are two temperatures, the one of the room
and of the cup, and there are two entropies. On top of that, the
timescale for heat exchange is large, allowing  maintenance of a spontanous  
difference in temperatures. A very similar situation occurs in a glass.
The fast modes (more precisely: fast modes combined with the bath)
play the role of the bath, while the slow configurational modes
play the role of the coffee. 

We may decompose the change $\d S$ of the total entropy $S=\S+\I$  
as the sum of the externally supplied entropy $d_e S=\dbarrm Q/T$ and the
internally produced entropy
\BEQ \d_i S= \frac{T-T_e}{T}\,\d\I \EEQ
It is positive because heat flows from high to low temperatures.
(At this point $T_e$ plays the role of a real second temperature!)
The entropy production per unit of time is $\d_i S/\d t$.

Within a set of smoothly related cooling experiments one is confined
to a surface $T_e=T_e(T,p)$, $p_e=p_e(T,p)$. 
This implies that $G$ has a discontinuous slope at the glass transition.
This difference in slope was first emphasized for spin
glasses~\cite{Nmaxmin}\cite{Ncomplexity}. Since its naive explanation
as being caused by a latent heat does not apply, since that has typically
not been observed, we were motivated to
find the physical background, 
which turned out to be the two-temperature approach,
discussed here. The difference in slopes implies for the Maxwell relation
\BEA\label{Maxwell=}
U^\prime &+&pV^\prime +T\dot V=(T_e-T\dot T_e)\I^\prime+T\dot \I T_e^\prime
\EEA
where we denoted $\p/\p_T|_p$ by a dot and $\p /\p p|_T$ by a prime.
In equilibrium ($T_e=T$) the right hand side vanishes, but
out of equilibrium there is no reason why it should. 
The second Ehrenfest relation is also modified, 
since it employs this Maxwell relation. It now becomes
\BEQ \label{modEf2} \frac{\Delta C_p}{TV}
=\Delta\alpha\frac{\d p}{\d T}
+\frac{1-\dot T_e}{V}\,\frac{\d \I}{\d T}
\EEQ
where $\d/\d T=\p/\p T+(\d p_g/d T)\p/\p p$ yields the total 
change along the glass transition line.
The last two terms are new, but would vanish in equilibrium.
The violation of the second Ehrenfest relation thus gives
information about the configurational entropy.

For atactic polystyrene Rehage and Oels
report a value $\Pi =1.09\approx 1$. Having realized that these authors
 used a short-time value for the compressibility in the glass,
we have taken the theoretical value from the first Ehrenfest relation, 
and  obtained $\Pi=0.77$~\cite{NEhren}. 
This can be described by eq. (\ref{modEf2}).
Values $\Pi<1$ also occur in the exact dynamical solution of a toy
model~\cite{Nhammer}.

In the past it was proven that $\Pi\ge 1$ is needed for mechanical
stability~\cite{DaviesJones}. The starting point was\
 the assumption that at the glass transition unspecified order 
parameters undergo a thermodynamic phase transition.
It was even shown that the same assumptions implied the
equality $\Pi=1$~\cite{Moynihan}. 
Since experiments mostly yielded $2<\Pi<5$, this remained a field
of confusion. It is now clear that
the assumption of a thermodynamic transition is false: the above
theoretical framework does  not involve it,
and the Rehage-Oels experiment has $\Pi< 1$.

A phase diagram  of amorphous ice
was presented recently~\cite{Mishima}. This approach, however,
employs equilibrium thermodynamics. In particular, it assumes
the validity of the Clausius-Clapeyron relation 
at the first order transition from low density amorphous ice 
to high density amorphous ice. However, for first order transitions to a
glassy state the Clausius-Clapeyron relation will be modified~\cite{NEhren}
\BEQ \label{CCmod}
{\Delta V}\frac{\d p_g}{\d T}=\frac{\Delta U+p\Delta V}{T}+
(1-\frac{\d T_e(T,p_g(T))}{\d T})\,\I_g
\EEQ
It would be interesting to test this for amorphous ice.

\subsection*{Apparent violations of the laws of thermodynamics}

The basis of our approach has been to show that glassy systems respect
the fundamental laws of thermodynamics.
However, if one ``forgets'' that the material is
in a non-equilibrium state, glass addresses
practically all thermodynamic paradoxes that
have been discussed in the past.

Take a piece of natural or volcanic glass and keep it at room
temperature for  extremely long time. The
material is in an excited, non-equilibrium state. In the course of
time it will slowly evolve towards the equilibrium crystalline state.
In doing so it will go to a state of lower energy and
 configurational entropy. As is well known, see e.g. ~\cite{DaviesJones},
 and described by eq. (\ref{2law}), this will generate heat.
There will be no true violation of the first and second law:
after the glass has reached equilibrium, such processes are impossible.

The released heat can be used in several ways:

\subsubsection*{ 
Energy out of ``nothing'': glass as a source of energy}

An efficient way to extract all configurational energy of
a glass is as follows.
One first brings it into its liquid state by heating;
then  one lets it carefully go into the crystalline state. 
This process allows to extract the configurational energy 
in a reasonable time.
 The energy needed for heating can in principle be
regained upon cooling. On top of that, one obtains the
crystallization heat $\Delta U$
(also called relaxation heat or latent heat),
corrected for the difference in specific heats of glass and crystal.

For metallic glasses this principle is well known to material scientist
~\cite{Greer}, and it equally holds for glasses and other amorphous 
materials, such as glassy volcanic rocks.
In analogy with metal alloys we can estimate the configurational energy
as some $10$ $kJ/{\it mole\,\, atoms}$. This is some 30 times
less than the conventional gain of energy from combustion of oil
and stone coal. This principle may also be applied to extract the
energy due to stresses in polycrystalline rocks.
Though these mechanisms are presently not attractive,
they may be relevant for travelling or living in outer space, or
perhaps for the mining of asteroids~\cite{Kowal}.

\subsubsection*{ Heat out of ``nothing'':
an isolated piece of glass heats itself slowly up}

When the glass is kept in an isolated container ($\dbarrm Q=0$),
the heat will be absorbed by the fast processes of the glass.
As they determine the temperature, the glass will slowly heat up.
Eq. (\ref{2law}) says that $T\dot\S+T_e\dot\I=0$. 
In the equilibration process $T_e$ will go down, but $T$,
the temperature that can be read off from a thermometer, will enhance.

An analogous manifestation is the fact that most of
the earth's interior is fluid due to radioactive decay.
Indeed, the radioactive content, with its decay time of
several billion years, is a slow and thus ``configurational'' process.
The generated heat is absorbed by the ``fast'' geological processes,
which maintain the fluid nature.

\subsubsection*{The quantum glass as an almost ideal motor}

A quantum glass in a bath at very low temperature, or a realistic 
glass in outer space with its 3 Kelvin background temperature,
can have a high effective temperature,
but basically have vanishing bath temperature.
The efficiency for transforming the dynamically released heat into work,
$\eta=1-T/T_e$, can be arbitrarily close to unity,
at the verge of violating the second law.

\subsubsection*{ Work out of ``nothing'':
a ``perpetuum mobile of the third kind''}

The heat released by the glass can also
be used to perform work by a suitable machine. We thus have a mobile
that performs work as long as the glass is not in equilibrium.
{}From geology it is known that, at least for Si-rich glasses,
this period may be astronomical~\cite{deJong}.
We may therefore call the non-equilibrium glass, combined with some
apparatus that performs the work, a ``perpetuum mobile of the third
kind''. Indeed, work $\d W=-\d Q$ is performed at the cost of
configurational entropy, which,  according to the third law,
should have been much closer to zero, in particular in cases where
the glass formation temperature is much larger than
 the room temperature.

It should  be admitted that for a small piece of glass
such a  mobile is merely an academic object.
In the above terrestrial analog, the heat released by radioactive decay
finally performs work by moving of the earth crust.

\subsubsection*{The third law and the residual entropy}

It is long known that on realistic time scales the zero temperature
entropy of glass does not vanish. It is equal to the confiurational
entropy of slow modes $\I$.

\subsubsection*{The zeroeth law and the effective temperature}

The configurational modes of a  window glass ``live'' at an 
effective temperature. During the glass formation process they  
got essentially stuck after the glass transition, which for window glass
occurs at some 1000 $K$. As they did not relax very much since then, 
the effective temperature $T_e$ of these modes is still 
close to, but below, the glass transition temperature.
It is thus not proper to say that a glass has a single, 
constant temperature: the one related to its slow modes is very
different from room temperature, and will be measured when
performing experiments on long enough timescales.

\section*{Appendix: The Ehrenfest relations}

At a first order phase transition from phase $A$ to phase $B$,
the slope of the transition line is given by the Clausius-Clapeyron
relation. It derives from continuity of the free enthalpy $G=U+pV-TS$.
The difference $\Delta G=G_A-G_B$ vanishes along the
transition line. 
Differentiating $\Delta G(T,p_g(T))=0$ one obtains
\BEQ  \label{CCrel}
\frac{\Delta U+p\Delta V}{T}={\Delta V}\frac{\d p_g}{\d T}
\EEQ

For second order phase transitions of classical type
one can differentiate $\Delta V=0$,
which yields the first Ehrenfest relation
\BEQ \label{Ehren1}
\Delta \alpha=\Delta \kappa\,\frac{\d p_g}{\d T}
\EEQ
It is  said to be violated in glasses, but we shall argue against that.

By considering $\Delta U=0$ one obtains the second Ehrenfest relation
\BEQ \label{orgEf2}
\frac{\Delta C_p}{TV}
=\Delta\alpha\frac{\d p_g}{\d T}
\EEQ
It is believed to be often satisfied for glasses, though not in all cases.
We shall show that in principle it is violated.
 
The information of both Ehrenfest relations is often coded in the
 so-called Prigogine-Defay ratio
\BEQ \Pi=\frac{\Delta C_p\Delta \kappa}{TV(\Delta\alpha)^2}\EEQ
Using eq. (\ref{Ehren1})  and (\ref{orgEf2}) it seen to
equal to unity in equilibrium. For glasses it is reported to lie typically
between 2 and 5.
We shall stress that also values below unity are possible, 
and actually already have been measured.

\end{multicols}

\acknowledgments
H. Bakker, H. Henrichs, B. de Jong, and G.W. Wegdam are acknowledged 
for discussion, and B. de Jong, M.H.G. Jacobs and L. Leuzzi
for critical reading of the manuscript.  
Part of this work was performed at the 
Max Planck Institute for Complex Systems in Dresden.

\begin{figure}[htb]
\label{rhoelsfig}
\epsfxsize=10cm
 \centerline{\epsffile{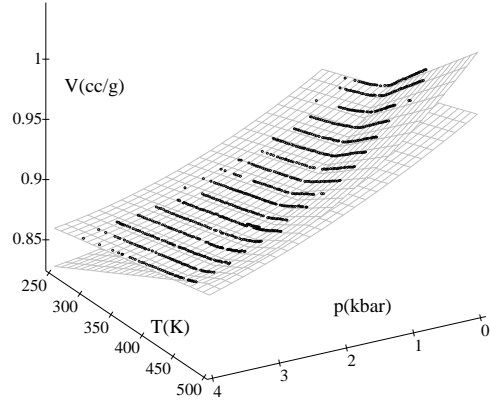}} 
\caption{Data of the glass transition for cooling
atactic polystyrene with 18 $K/h$, taken
from Rehage and Oels (1976):
specific volume $V$ ($cm^3/g$) versus temperature $T$
($K$) at various pressures $p$ ($k\,bar$). The data in the liquid
essentially lie on a smooth surface, and so do the data in the glass.
The first Ehrenfest relation describes no more than the
intersection of these surfaces, and is thus automatically satisfied.
 The values for the compressibility that follow from these data 
 will generally differ from results obtained via other experiments.}
\end{figure}
\end{document}